\documentclass[a4paper,11pt]{article}
\usepackage{pos}
\usepackage{subcaption}
\usepackage{siunitx}
\usepackage{enumitem}

\title{Validation of Electromagnetic Showers in CORSIKA 8}
 \ShortTitle{Electromagnetic shower validation in C8}

\author*[a]{Alexander Sandrock}
\author[b]{Jean-Marco Alameddine}
\author[c]{Felix Riehn}

\affiliation[a]{Bergische Universität Wuppertal,
  Fakultät für Mathematik und Naturwissenschaften,\\
  Gaußstraße 20, 42119 Wuppertal, Germany}

\affiliation[b]{Technische Universität Dortmund, Fakultät Physik,\\
  Otto-Hahn-Straße 4a, 44227 Dortmund, Germany}

\affiliation[c]{Instituto Galego de Física de Altas Enerxías (IGFAE),
  Universidade de Santiago de Compostela,\\
  Rúa de Xoaquín Díaz de Rábago s/n, 15705 Santiago de Compostela, Galicia, Spain}

\onbehalf{for the CORSIKA~8 collaboration} 

\emailAdd{asandrock@icecube.wisc.edu}
\emailAdd{jean-marco.alameddine@tu-dortmund.de}
\emailAdd{friehn@lip.pt}

\abstract{The air shower simulation code CORSIKA has served as a key
part of the simulation chain for numerous astroparticle physics
experiments over the past decades. Due to retirement of the original
developers and the increasingly difficult maintenance of the monolithic
Fortran code of CORSIKA, a new air shower simulation framework has been
developed over the course of the last years in C++, called CORSIKA~8.
Besides the hadronic and muonic component, the electromagnetic component
is one of the key constituents of an air shower. The cascade producing
the electromagnetic component of an air shower is driven by
bremsstrahlung and photoproduction of electron-positron pairs. At
ultrahigh energies or in media with high densities, the bremsstrahlung and pair
production processes are suppressed by the Landau-Pomeranchuk-Migdal
(LPM) effect, which leads to more elongated showers compared to showers
without the LPM suppression. Furthermore, photons at higher energies can
produce muon pairs or interact hadronically with nucleons in the target
medium, producing a muon component in electromagnetic air showers.
In this contribution, we compare electromagnetic showers simulated with
the latest Fortran version of CORSIKA and CORSIKA~8, which uses the library
PROPOSAL for the electromagnetic component. While earlier
validations of CORSIKA~8 electromagnetic showers focused on showers of
lower energy, the recent implementation of the LPM effect, photo pair
production of muons, and of photohadronic interactions allows now to
make a physics-complete comparison also at high energies.}

\ConferenceLogo{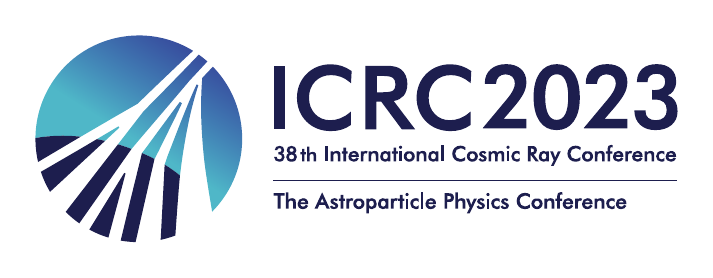}

\FullConference{%
38th International Cosmic Ray Conference (ICRC2023)\\
  26 July - 3 August, 2023\\
  Nagoya, Japan}

\begin{document}
\maketitle
\section{Introduction}
Many astroparticle physics experiments have relied on CORSIKA
\cite{Heck98CORSIKA} for the simulation of extensive air showers. The
Fortran version, originally developed for the KASCADE experiment, is
currently at version 7.7500 and still has an excellent performance. However,
its monolithic design and hand-optimized code make its further development
and the addition of new features increasingly difficult. To overcome these
limitations, a new modular simulation framework for extensive air showers called
CORSIKA~8 has been developed in C++ over the course of the last years (cf. also
\cite{huege_icrc2023}).

The electromagnetic component of air showers is formed as the result of a
cascade of bremsstrahlung and pair production processes. In the Fortran
versions of CORSIKA (C7 for short) the processes responsible for this shower
component were simulated by a customized version of EGS4 \cite{EGS4}, which is
deeply integrated in the CORSIKA source code. In its place, in CORSIKA~8 (C8)
this task is carried out using the particle propagation library PROPOSAL
\cite{Koehne13PROPOSAL,Dunsch18PROPOSAL,Alameddine20PROPOSAL,Alameddine2023}, a modular
C++14 library with Python bindings that can be used for the propagation of
electrons, positrons, muons, tau leptons as well as photons. The current version of PROPOSAL is 7.6.2.

\begin{figure}
  \begin{subfigure}{0.495\textwidth}
    \includegraphics[width=\textwidth]{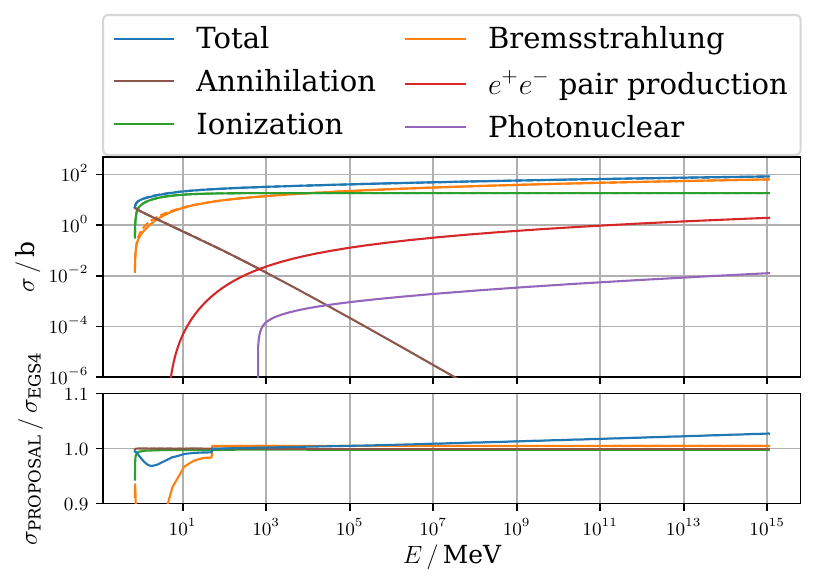}
    \caption{Positron cross-sections compared between C7 and C8.}
  \end{subfigure}
  \begin{subfigure}{0.495\textwidth}
    \includegraphics[width=\textwidth]{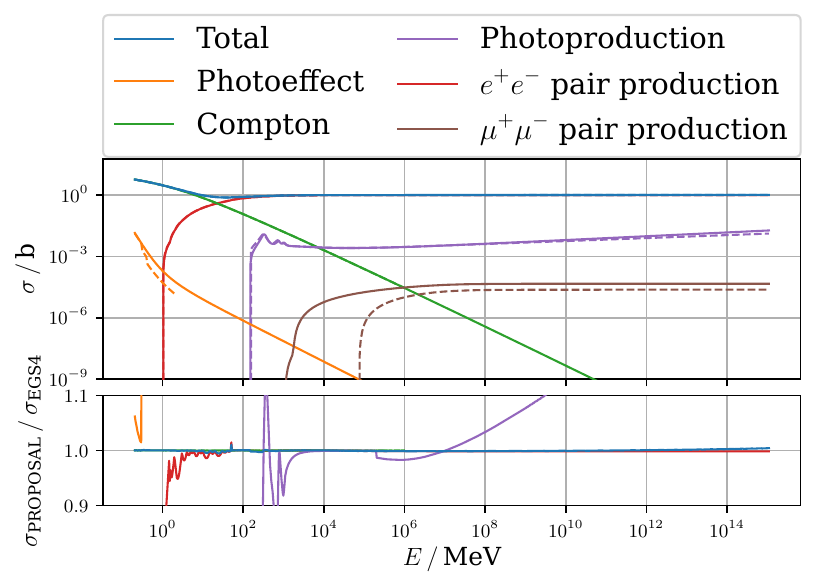}
    \caption{Photon cross-sections compared between C7 and C8.}
  \end{subfigure}
  \caption{Total stochastic cross-sections for positrons and photons in air. The
    losses smaller than \SI{0.2}{MeV} are treated continuously to obtain a
    finite cross-section. The dashed lines refer to cross-sections from C7, the solid lines show the implementation in C8.} 
  \label{fig:xsec}
\end{figure}

PROPOSAL offers
several parametrizations for the cross-sections of the pertinent processes;
the default choice for C8 differs only in a few minor details from the
parametrizations used in C7:
 Rayleigh scattering is not implemented in CORSIKA~8,
 triplet production $e^\pm \rightarrow e^\pm e^+ e^-$ is not implemented in
    CORSIKA~7, and
   the parametrization of the photoelectric effect and the calculation
    of secondary particles in photohadronic interactions are different.
The cross-sections implemented differ at most by a few percent, and that only
at the lowest or extremely high energies or in regions where the contribution
of the process is very small (cf. Fig.~\ref{fig:xsec} for exemplary comparisons
of positron and photon cross-sections between C7 and C8).

\section{Recent improvements to the electromagnetic shower simulation}
Since the last ICRC, we have implemented several enhancements in C8. The most
important ones are the treatment of photohadronic interactions, the
photo-pairproduction of muons, and the implementation of the Landau-Pomeranchuk%
-Migdal (LPM) effect. Furthermore, the new improved implementation of the Moli\`ere multiple scattering in PROPOSAL allowed to use this more precise treatment instead of the faster approximation by Highland used earlier in C8 for speed reasons.

Photohadronic interactions are a subdominant interaction process of high-energy
photons, since the cross-section is about 1\% of the pair production process.
However, the decay products of the produced hadrons constitute the dominant
amount of the muon content in electromagnetic showers and a non-negligible
portion in hadronic showers via electromagnetic sub-showers \cite{mueller_muon}. In C7, the
photohadronic interaction was treated with routines for single-pion or two-pion production at resonances and the hadron dual-parton model at low
energies below \SI{80}{GeV}, and with the chosen high-energy hadronic
interaction model at higher energies. In C8, the low-energy portion is treated
by the code SOPHIA \cite{SOPHIA}, while at higher energies the high-energy
hadronic interaction model is called as well. Nuclear effects in the photohadronic interaction are neglected in both C7 and C8, apart from a shadowing correction to the total cross-section.

The remaining part of the muon content in electromagnetic air showers is due to
photopair production of muons by photons. As in C7, the cross section from
\cite{MuonPair} has been implemented with an additional term in analogy to
\cite{atomic_electrons} to take into account the production of muon pairs on
atomic electrons.

\begin{figure}
  \begin{subfigure}{0.485\textwidth}
      \includegraphics[width=\textwidth]{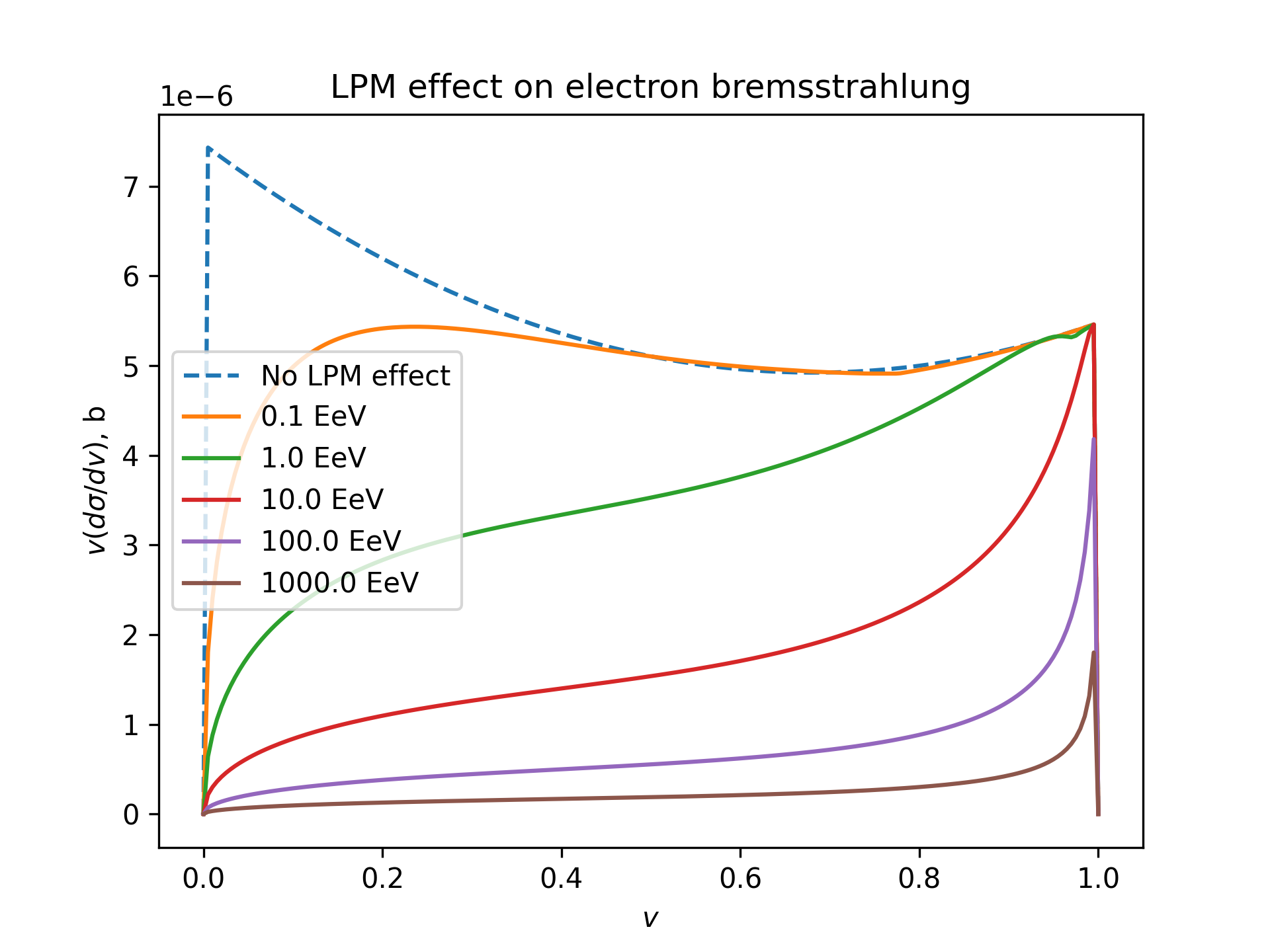}
      \caption{Bremsstrahlung cross section in air with and without the LPM effect.}
  \end{subfigure}
  \begin{subfigure}{0.485\textwidth}
      \includegraphics[width=\textwidth]{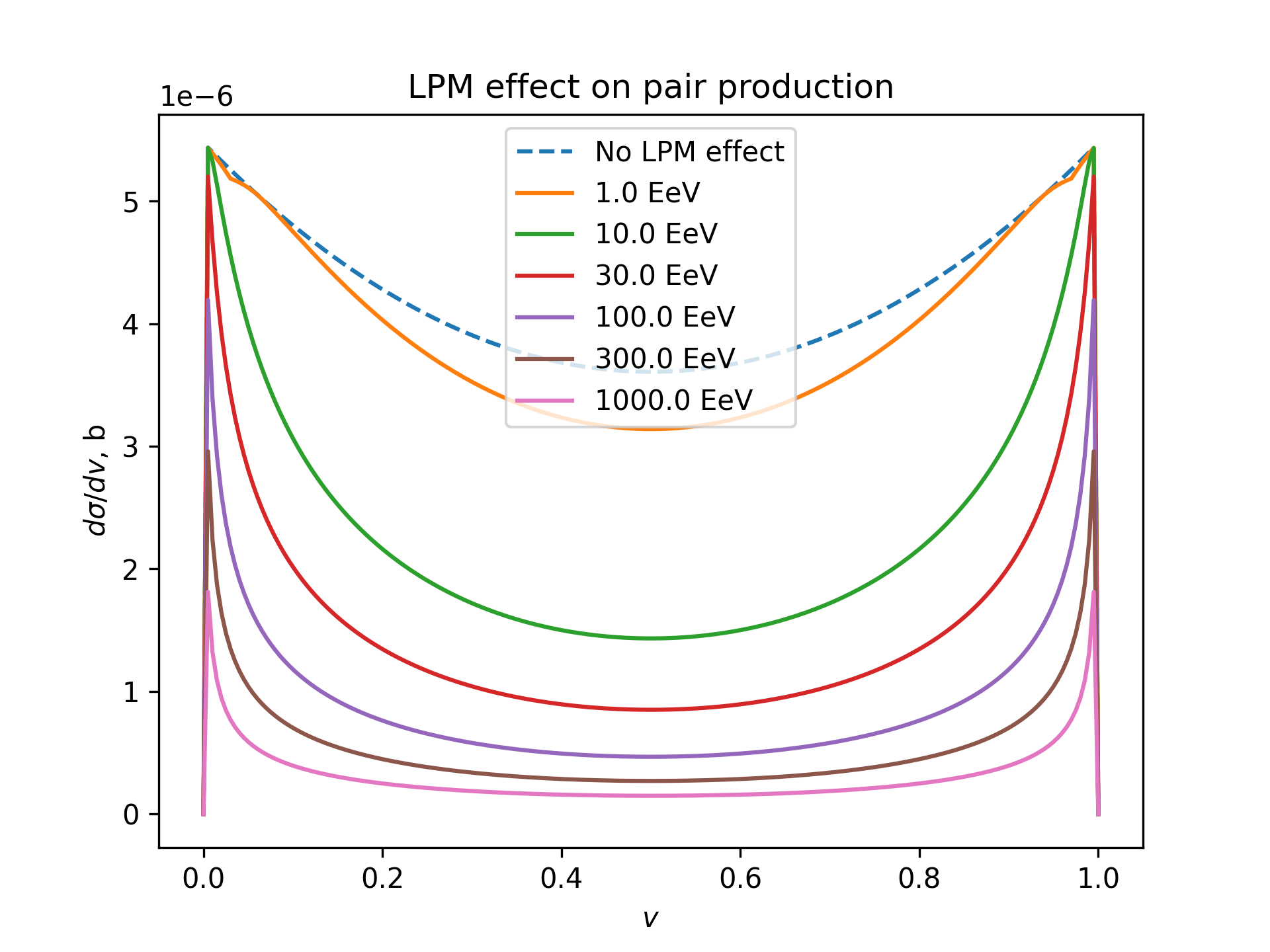}
      \caption{Pair production cross section in air with and without the LPM effect.}
  \end{subfigure}
  \caption{Differential cross-sections of bremsstrahlung and pair production
  at very high energies in air at sea level pressure to illustrate the LPM
  effect. $v$ denotes the fraction of the primary particle energy going into the secondary photon or positron, respectively.}
  \label{fig:lpm_xsec}
\end{figure}

The LPM effect is a density-dependent effect which suppresses the emission of
bremsstrahlung photons with energies small compared to the electron energy, as
well as the production of pairs with a roughly equal distribution of energy
between electron and positron at high energies (cf. Fig.~\ref{fig:lpm_xsec}).
In the Monte Carlo particle shower simulation, the LPM suppression is treated with a Neumann rejection
method: after an interaction has been sampled, an additional random number is
drawn and if the random number is larger than the ratio between the cross-%
section with and without LPM suppression, the interaction is rejected (for the
description in C7 see \cite{C7_LPM}).

\section{Validation of electromagnetic showers}
To validate the simulation of electromagnetic showers in C8, we compare against C7 showers
with an initial electron at various primary energies and particle cuts as a widely used reference. The
atmosphere is the US standard atmosphere, the chosen low- and high-energy
hadronic interaction models are FLUKA \cite{fluka} and \textsc{Sibyll} 2.3d \cite{sibyll}, respectively. We compare the
longitudinal profiles of particle numbers and charge excess, as well as the
lateral distributions and energy spectra of the showers near the shower maximum.

\subsection{100~TeV showers}
\begin{figure}
  \begin{subfigure}{0.485\textwidth}
    \includegraphics[width=\textwidth]{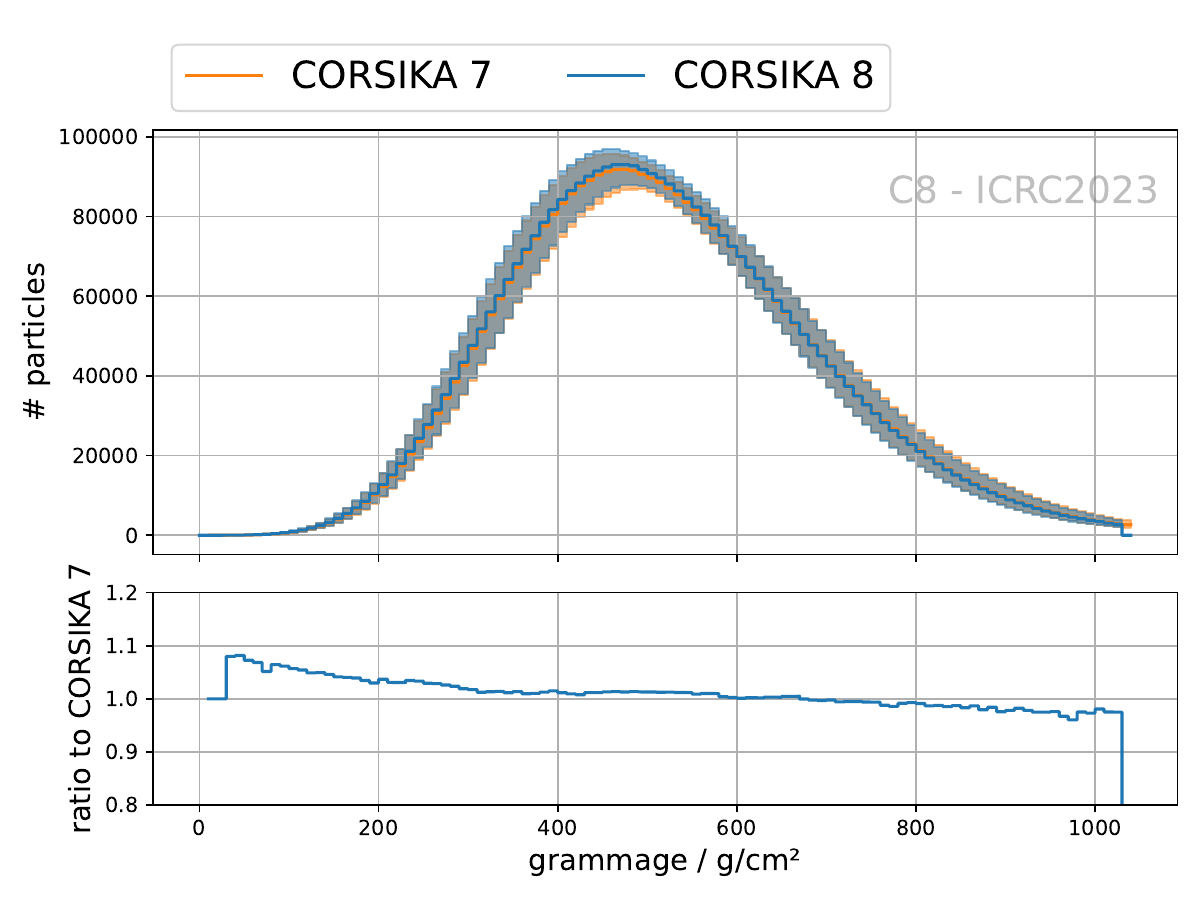}
    \caption{Longitudinal profile of charged particles.}
    \label{fig:100TeV_1MeV_char_long}
  \end{subfigure}
  \begin{subfigure}{0.485\textwidth}
    \includegraphics[width=\textwidth]{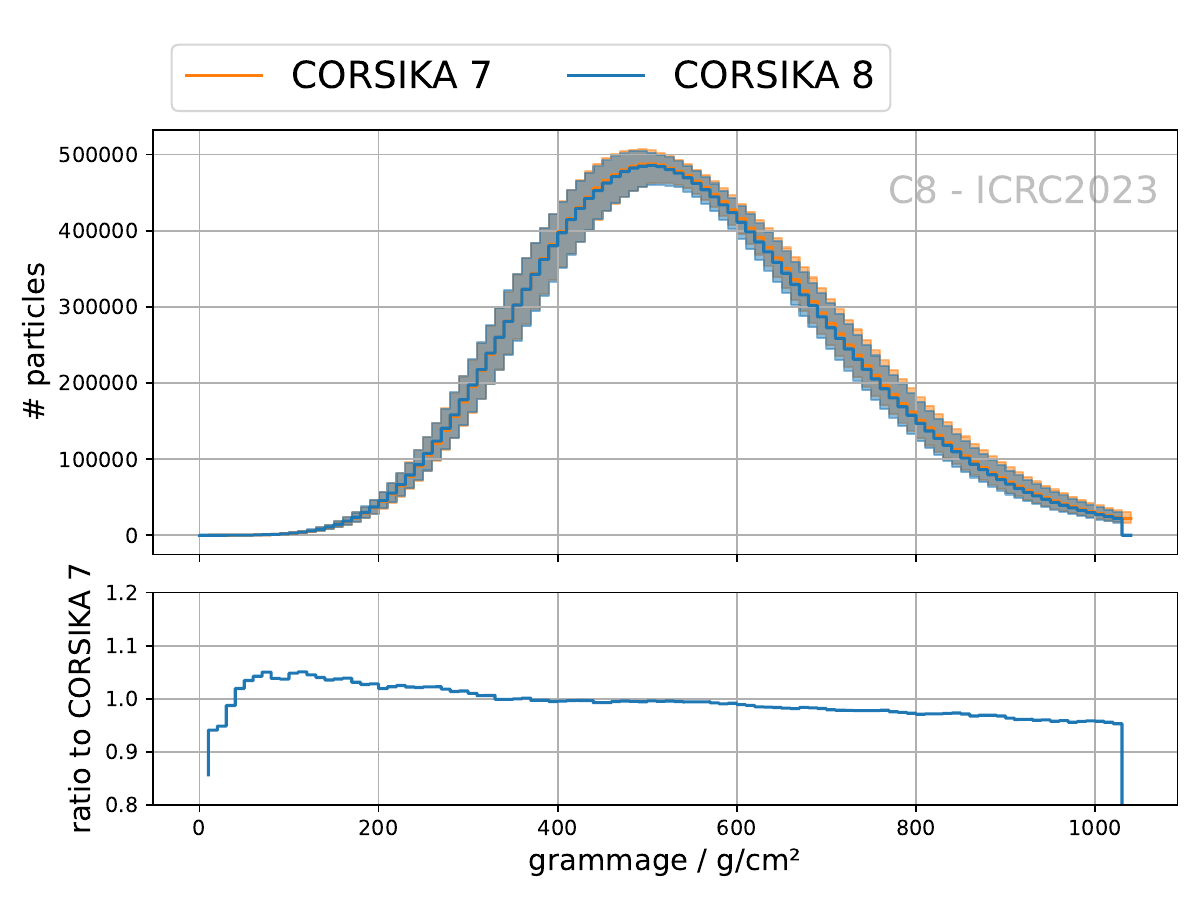}
    \caption{Longitudinal profile of photons.}
    \label{fig:100TeV_1MeV_phot_long}
  \end{subfigure} \\
  \begin{subfigure}{0.485\textwidth}
    \includegraphics[width=\textwidth]{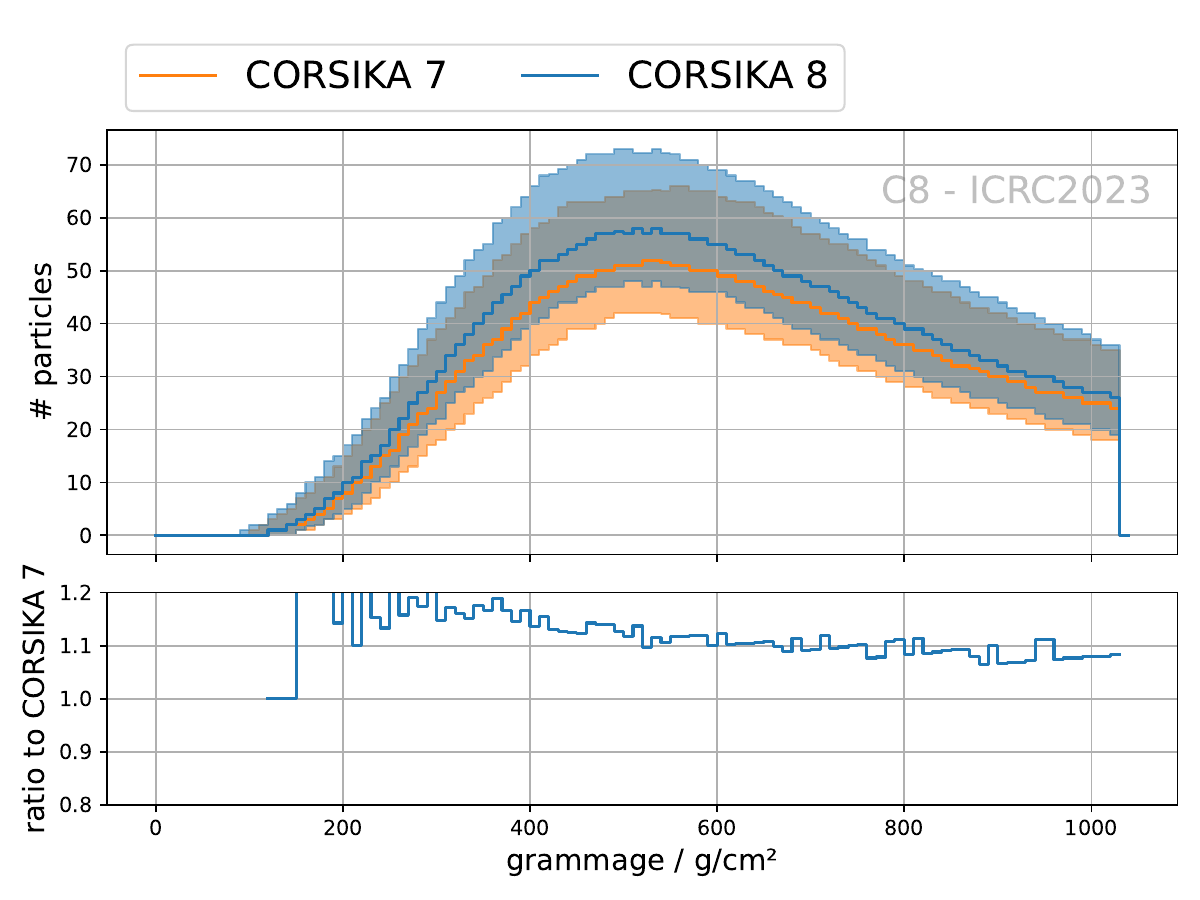}
    \caption{Longitudinal profile of muons.}
    \label{fig:100TeV_1MeV_mu_long}
  \end{subfigure}
  \begin{subfigure}{0.485\textwidth}
    \includegraphics[width=\textwidth]{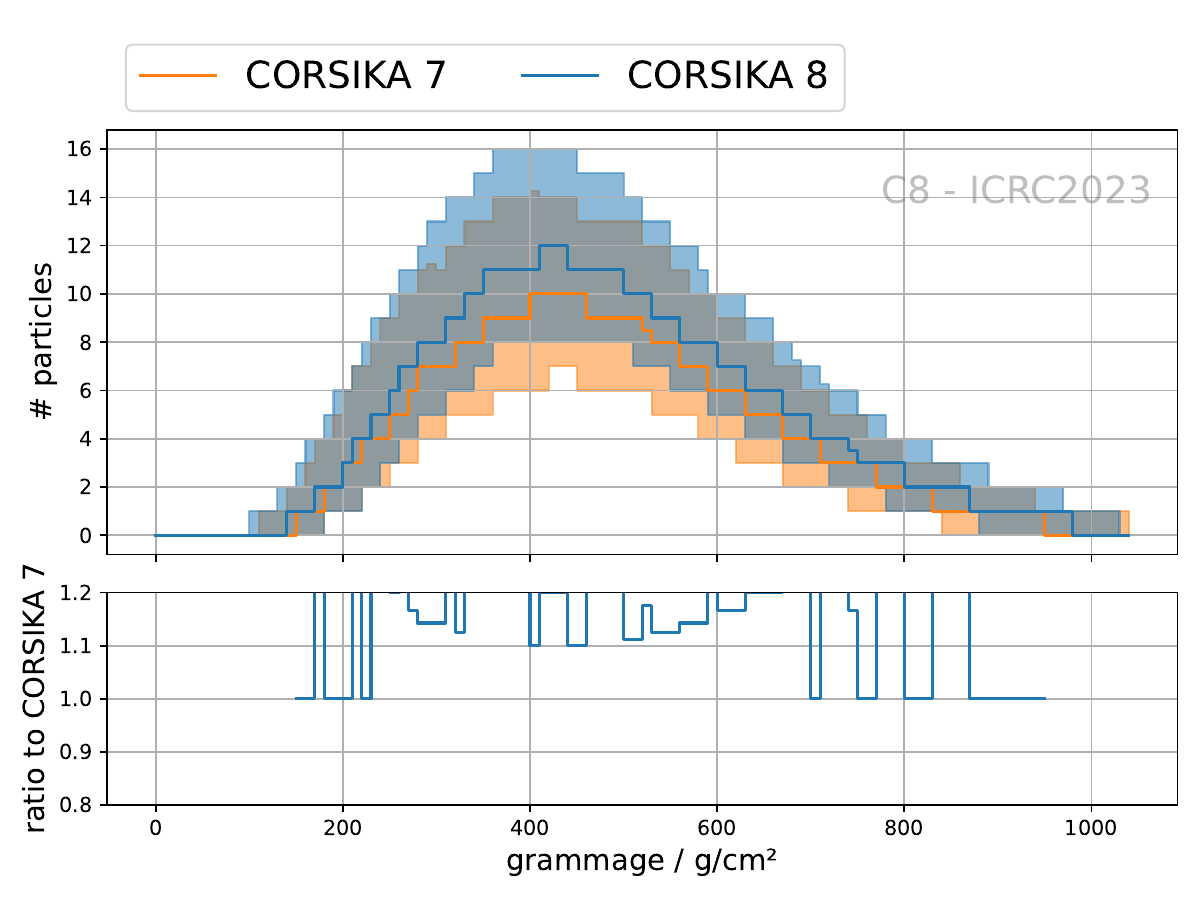}
    \caption{Longitudinal profile of hadrons.}
    \label{fig:100TeV_1MeV_had_long}
  \end{subfigure} \\
  \begin{subfigure}{0.485\textwidth}
      \includegraphics[width=\textwidth]{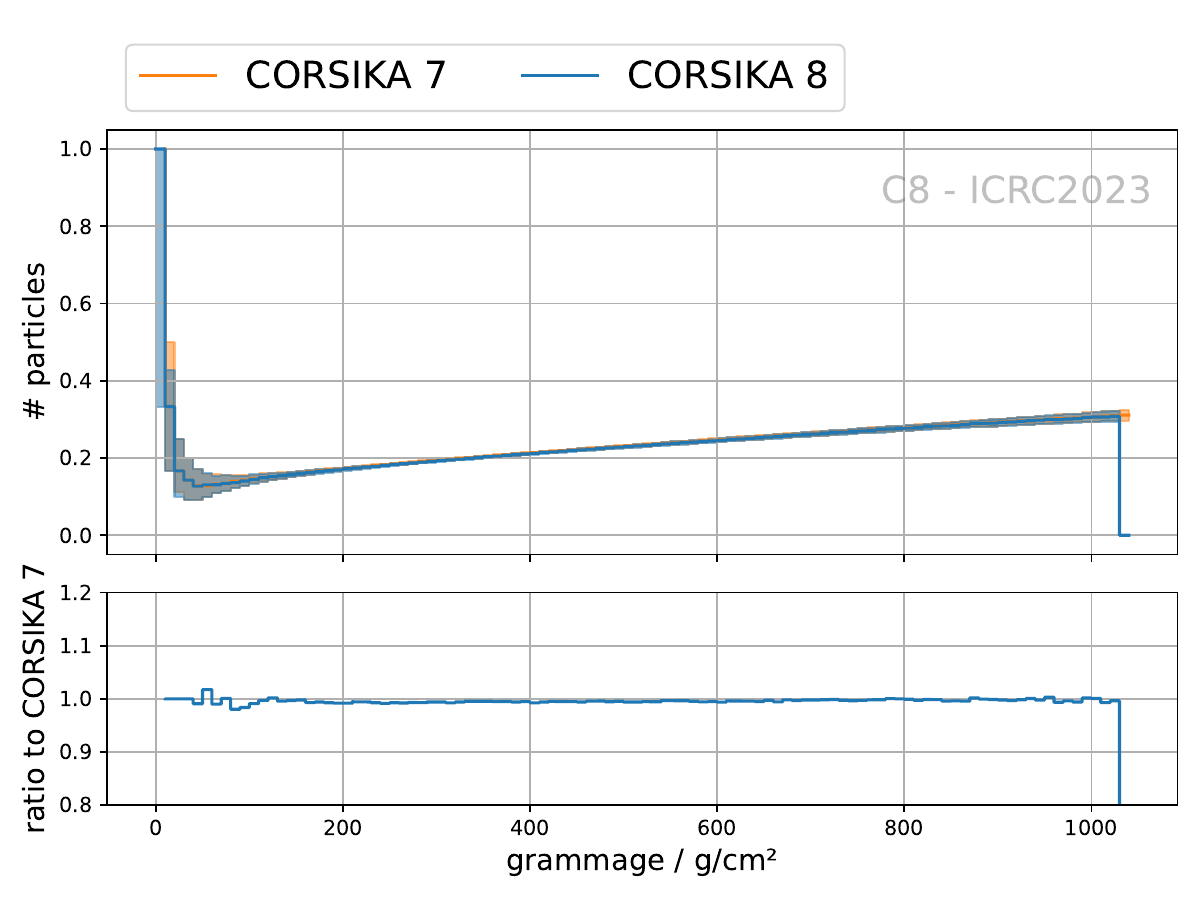}
      \caption{Longitudinal profile of the charge excess.}
      \label{fig:100TeV_1MeV_char_exc}
  \end{subfigure}
  \begin{subfigure}{0.485\textwidth}
      \includegraphics[width=\textwidth]{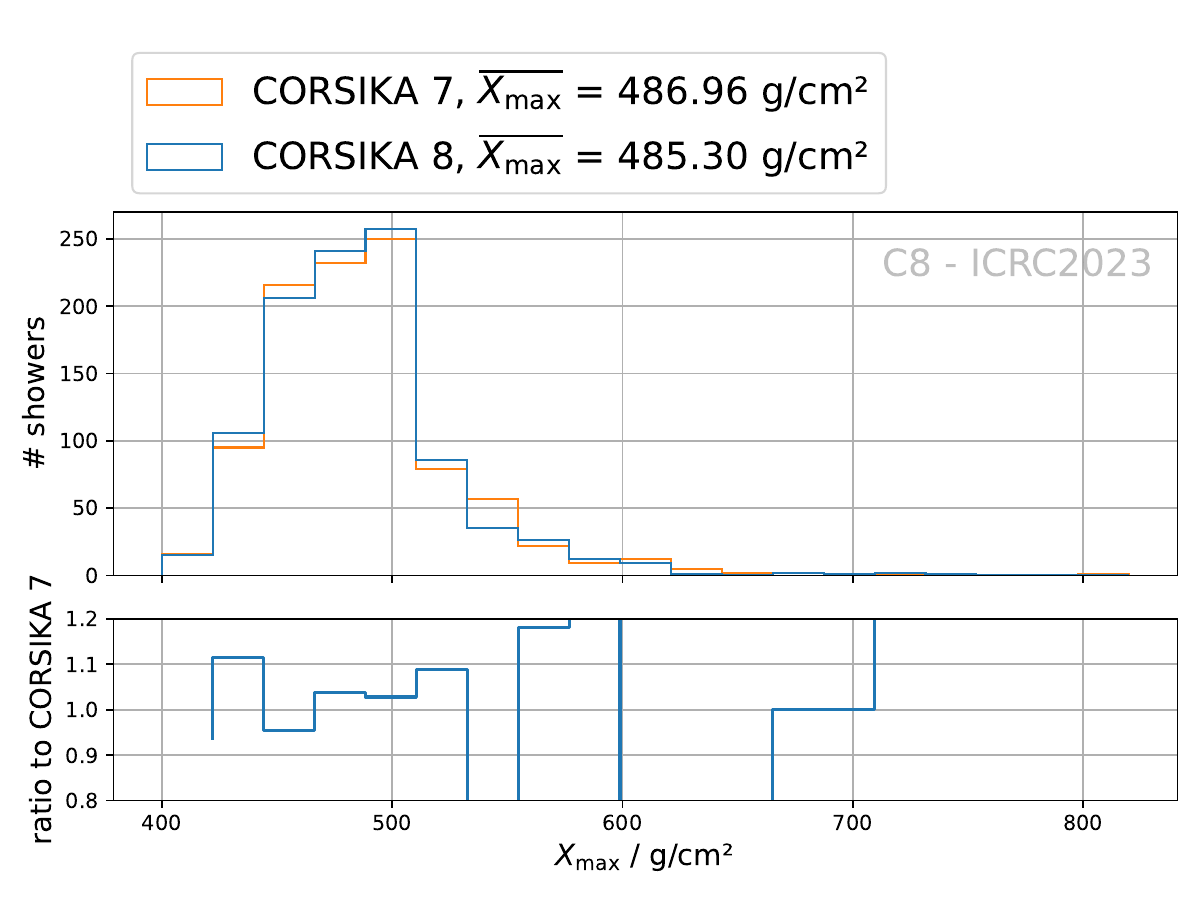}
      \caption{Distribution of shower maxima.}
      \label{fig:100TeV_1MeV_Xmax}
  \end{subfigure}
  \caption{Longitudinal profiles and $X_\text{max}$ distribution for \SI{100}{TeV} showers with \SI{1}{MeV} particle cut}
  \label{fig:100TeV_1MeV}
\end{figure}

In Fig.~\ref{fig:100TeV_1MeV}, we show the longitudinal profiles of 1000 showers with an \SI{100}{TeV} $e^-$ as primary particle and \SI{1}{MeV} particle cut energy. The charged particles and photon longitudinal profiles agree to within better than 5\%, and the charge excess to better than 3\%; the longitudinal profiles of muons and hadrons show about 10\% more particles, and the shower maxima are earlier by about \SI{2}{g/cm^2}. The shaded area around the profiles denotes the inter-quartile range on the histograms.

\begin{figure}
    \begin{subfigure}{0.485\textwidth}
        \includegraphics[width=\textwidth]{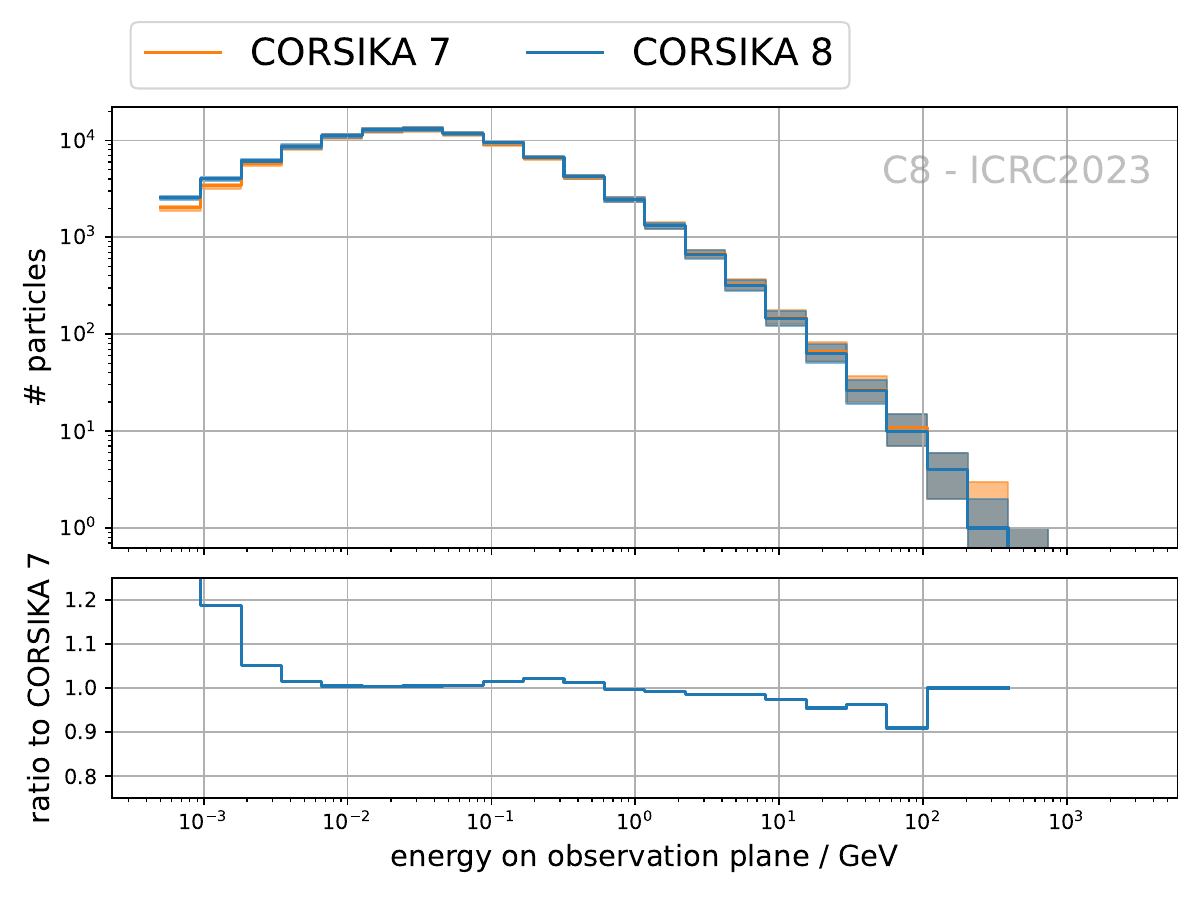}
        \caption{Energy spectrum of charged particles.}
    \end{subfigure}
    \begin{subfigure}{0.485\textwidth}
        \includegraphics[width=\textwidth]{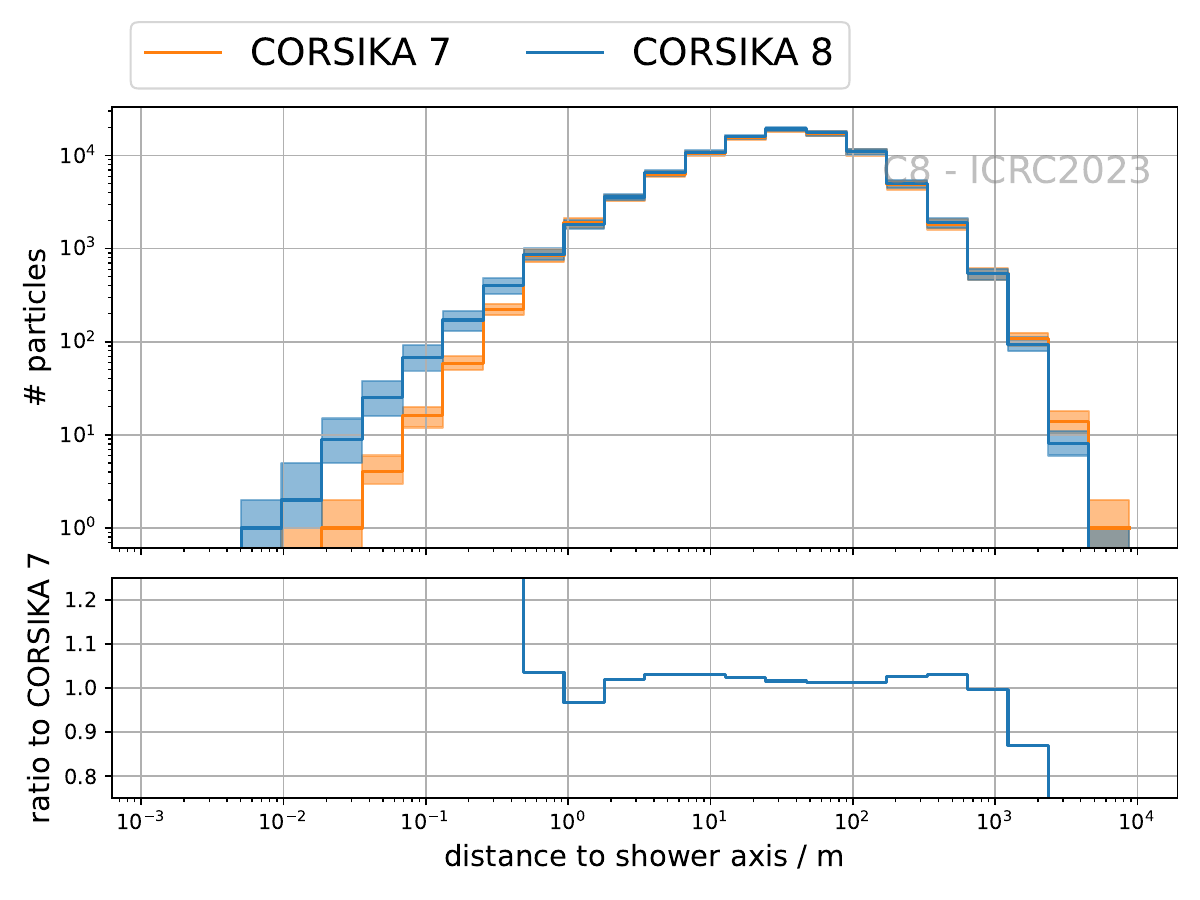}
        \caption{Lateral distributions of charged particles.}
    \end{subfigure} \\
    \begin{subfigure}{0.485\textwidth}
        \includegraphics[width=\textwidth]{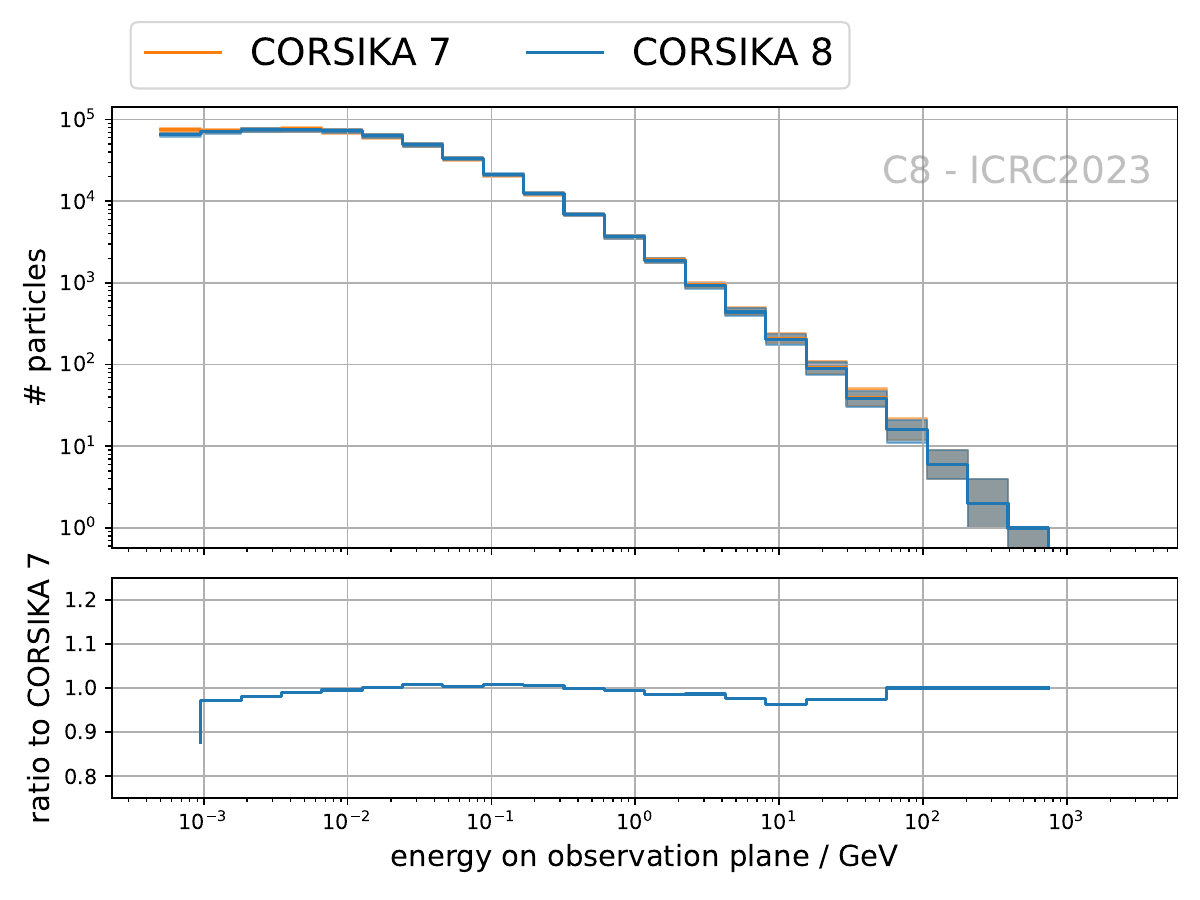}
        \caption{Energy spectrum of photons.}
    \end{subfigure}
    \begin{subfigure}{0.485\textwidth}
        \includegraphics[width=\textwidth]{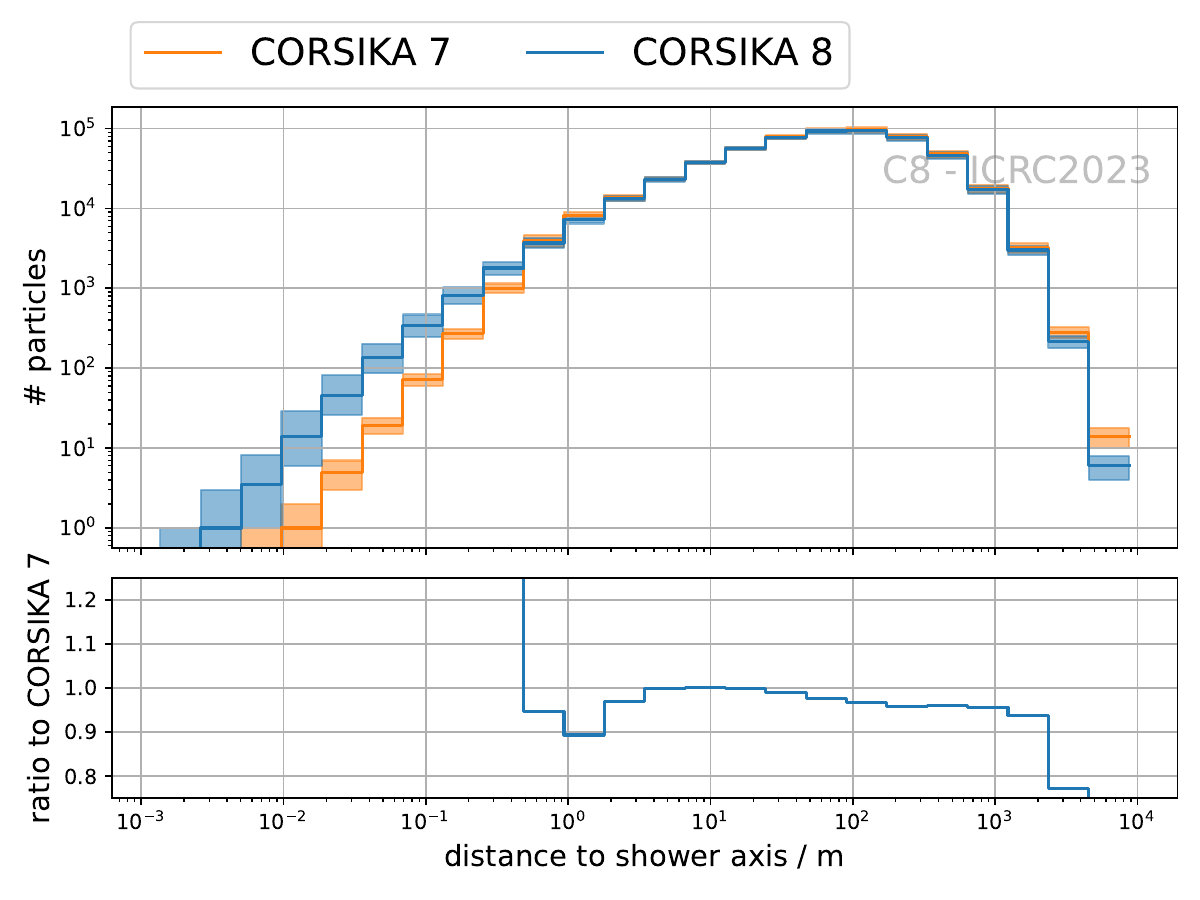}
        \caption{Lateral distributions of photons.}
    \end{subfigure}
    \caption{Energy spectra and lateral distributions of \SI{100}{TeV} showers near the shower maximum.}
    \label{fig:100TeV_Xmax}
\end{figure}

In Fig.~\ref{fig:100TeV_Xmax}, we show energy spectra and lateral distributions of particles near the shower maximum. Overall a good agreement is observed, except in the lowest and highest bins the differences amount to no more than 5\%; the lateral distribution functions agree within 3\% except for the bins nearer than about \SI{1}{m} from the shower core.

\begin{figure}
    \begin{subfigure}{0.49\textwidth}
        \includegraphics[width=\textwidth]{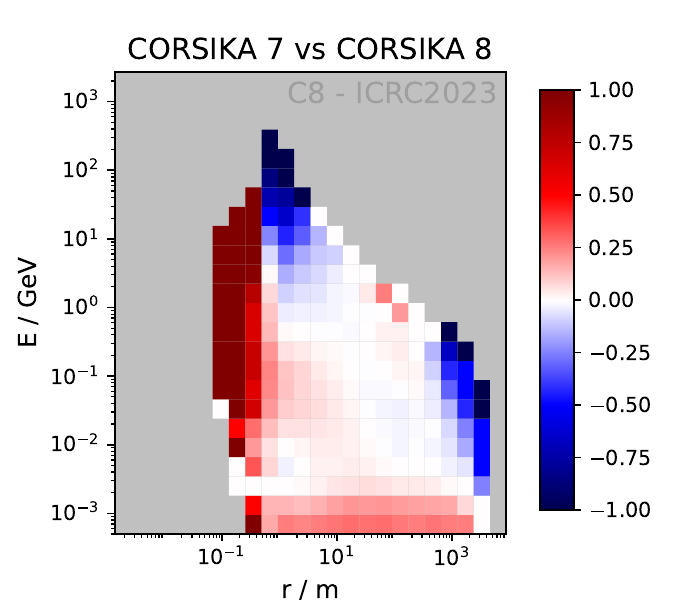}
        \caption{Charged particles.}
    \end{subfigure}
    \begin{subfigure}{0.49\textwidth}
        \includegraphics[width=\textwidth]{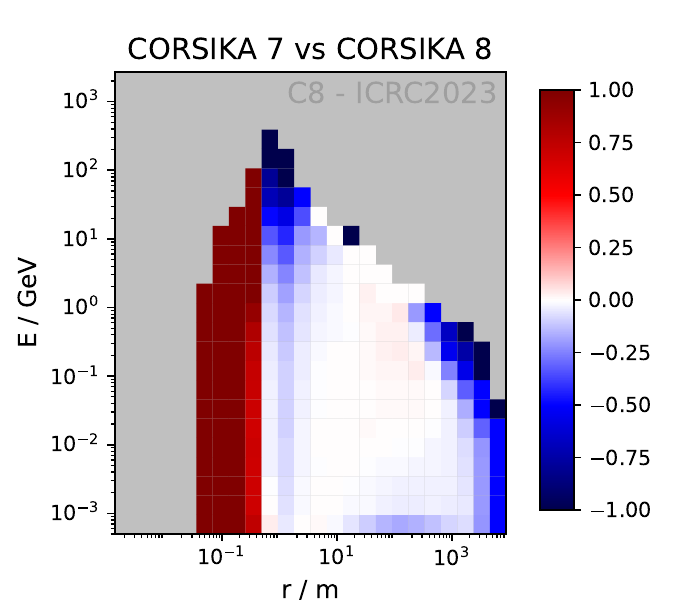}
        \caption{Photons.}
    \end{subfigure}
    \caption{Deviation from unity of the ratio of two-dimensional distributions in energy and radius near the shower maximum.}
    \label{fig:100TeV_2D}
\end{figure}

The ratio between the two-dimensional distributions in radius and energy near the shower maximum of these showers for charged particles and photons with C7 and C8 are shown in \ref{fig:100TeV_2D}. These distributions show some small differences in low-energy charged particles in addition to the disagreement between C7 and C8 at small radii visible in the one-dimensional lateral distributions.

\subsection{100~EeV showers and the LPM effect}
\begin{figure}
    \begin{subfigure}{0.485\textwidth}
        \includegraphics[width=\textwidth]{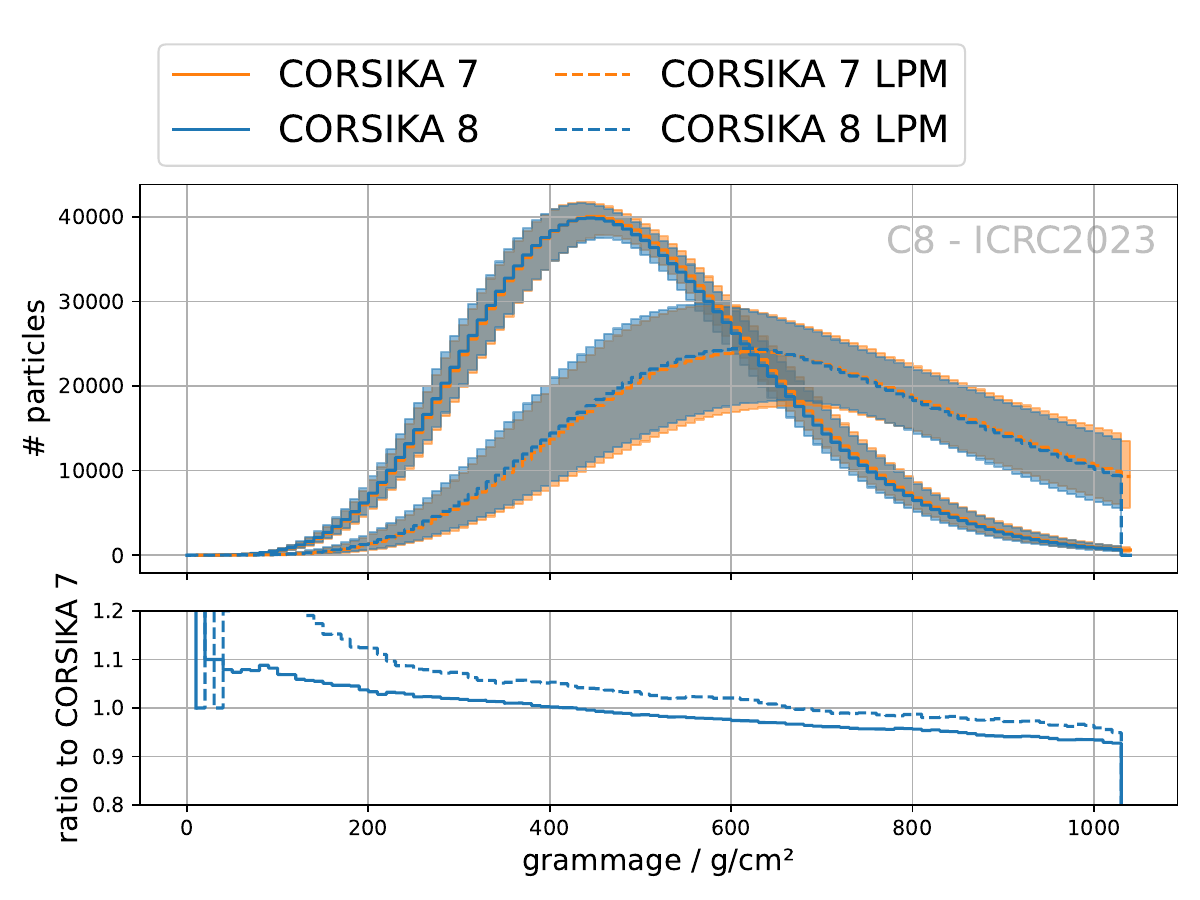}
        \caption{Longitudinal profile of charged particles.}
    \end{subfigure}
    \begin{subfigure}{0.485\textwidth}
        \includegraphics[width=\textwidth]{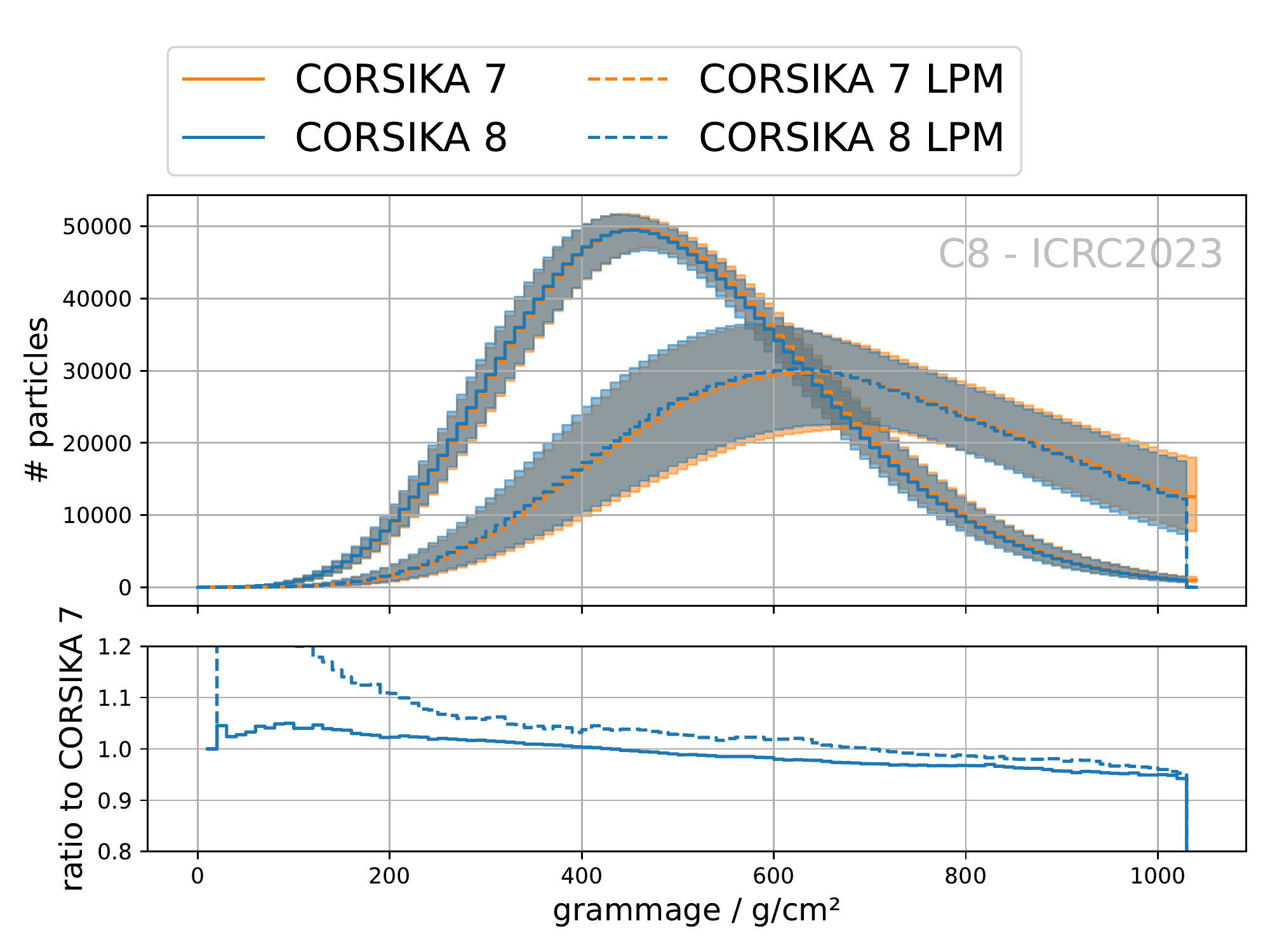}
        \caption{Longitudinal profile of photons.}
    \end{subfigure}
    \caption{Longitudinal profiles for \SI{100}{EeV} electromagnetic showers with a particle cut of \SI{100}{TeV}. The showers without the LPM effect are shown in solid lines, while the showers with LPM Effect are shown in dashed lines.}
    \label{fig:100EeV_LPM}
\end{figure}

In Fig.~\ref{fig:100EeV_LPM}, the longitudinal profiles of charged particles and photons for 5000 electromagnetic showers with a primary energy of \SI{100}{EeV} and a particle cut of \SI{100}{TeV} are shown with and without the LPM effect option. This high particle cut was chosen for performance reasons, as only the highest energy interactions are expected to be influenced sufficiently strongly by the LPM effect to change the global behaviour of the showers. As expected, LPM showers develop slower and reach their maximum later. The agreement is not as good as at the lower energies investigated above, but the deviations in the profiles are not worse than 5--10\%. The somewhat larger number of charged particles in C8 could be due to the increasing role of triplet pair production at these extremely high energies, since this process is not taken in to account in C7.
\section{Conclusions}
In this contribution, the electromagnetic shower simulation in CORSIKA~8 was validated by comparison to simulated showers from CORSIKA~7. The implementation of photohadronic interactions, photo pair production of muons, and the Landau-Pomeranchuk-Migdal effect allowed to extend earlier comparisons at lower energies \cite{Alameddine21_ICRC} to higher energies. The charge excess and longitudinal profiles agree to better than 3--5\%, the energy spectra and lateral distribution functions near the shower maximum show good agreement except very near the shower core in a logarithmic plot, and the longitudinal profiles of LPM showers agree within not worse than about 10\%.

Some disagreement between C7 and C8 is to be expected, since the code bases are disjunct except for the externally provided hadronic interaction models, and some rarer processes are treated in only one of the codes. Since the electromagnetic physics implemented is the essentially same, these comparisons also serve as a validation of C7. Overall, the electromagnetic showers simulated in C8 agree well with C7, including both the electrons, positrons, and photons as well as muons and hadrons not studied in earlier validations.
\acknowledgments
This research has been funded by the Deutsche Forschungsgemeinschaft (DFG) and the Lamarr institute.
\bibliography{corsika}

\clearpage
\section*{Full Authors List: CORSIKA~8 Collaboration}

%
\scriptsize
\noindent
J.M.~Alameddine$^{1}$,
J.~Albrecht$^{1}$,
J.~Alvarez-Mu\~niz$^{2}$,
J.~Ammerman-Yebra$^{2}$,
L.~Arrabito$^{3}$,
J.~Augscheller$^{4}$,
A.A.~Alves Jr.$^{4}$,
D.~Baack$^{1}$,
K.~Bernl\"ohr$^{5}$,
M.~Bleicher$^{6}$,
A.~Coleman$^{7}$,
H.~Dembinski$^{1}$,
D.~Els\"asser$^{1}$,
R.~Engel$^{4}$,
A.~Ferrari$^{4}$,
C.~Gaudu$^{8}$,
C.~Glaser$^{7}$,
D.~Heck$^{4}$,
F.~Hu$^{9}$,
T.~Huege$^{4,10}$,
K.H.~Kampert$^{8}$,
N.~Karastathis$^{4}$,
U.A.~Latif$^{10}$,
H.~Mei$^{11}$,
L.~Nellen$^{12}$,
T.~Pierog$^{4}$,
R.~Prechelt$^{13}$,
M.~Reininghaus$^{4}$,
W.~Rhode$^{1}$,
F.~Riehn$^{14,2}$,
M.~Sackel$^{1}$,
P.~Sala$^{15}$,
P.~Sampathkumar$^{4}$,
A.~Sandrock$^{8}$,
J.~Soedingrekso$^{1}$,
R.~Ulrich$^{4}$,
D.~Xu$^{11}$,
E.~Zas$^{2}$
\\
\begin{description}[labelsep=0.2em,align=right,labelwidth=0.7em,labelindent=0em,leftmargin=2em,noitemsep]
\item[$^{1}$] Technische Universit\"at Dortmund (TU), Department of Physics, Dortmund, Germany
\item[$^{2}$] Universidade de Santiago de Compostela, Instituto Galego de F\'\i{}sica de Altas Enerx\'\i{}as (IGFAE), Santiago de Compostela, Spain
\item[$^{3}$] Laboratoire Univers et Particules de Montpellier, Universit\'e de Montpellier, Montpellier, France
\item[$^{4}$] Karlsruhe Institute of Technology (KIT), Institute for Astroparticle Physics (IAP), Karlsruhe, Germany
\item[$^{5}$] Max Planck Institute for Nuclear Physics (MPIK), Heidelberg, Germany
\item[$^{6}$] Goethe-Universit\"at Frankfurt am Main, Institut f\"ur Theoretische Physik, Frankfurt am Main, Germany
\item[$^{7}$] Uppsala University, Department of Physics and Astronomy, Uppsala, Sweden
\item[$^{8}$] Bergische Universit\"at Wuppertal, Department of Physics, Wuppertal, Germany
\item[$^{9}$] Peking University (PKU), School of Physics, Beijing, China
\item[$^{10}$] Vrije Universiteit Brussels, Department of Physics and Astrophysics (DNTK), Brussels, Belgium
\item[$^{11}$] Tsung-Dao Lee Institute (TDLI), Shanghai Jiao Tong University, Shanghai, China
\item[$^{12}$] Universidad Nacional Aut\'onoma de M\'exico (UNAM), Instituto de Ciencias Nucleares, M\'exico, D.F., M\'exico
\item[$^{13}$] University of Hawai'i at Manoa, Department of Physics and Astronomy, Honolulu, USA
\item[$^{14}$] Laborat\'orio de Instrumenta\c{c}\~ao e F\'\i{}sica Experimental de Part\'\i{}culas (LIP), Lisboa, Portugal
\item[$^{15}$] Fluka collaboration
\end{description}

%

\end{document}